\documentclass[twocolumn,showpacs,preprintnumbers,amsmath,amssymb,prb]{revtex4}

\usepackage{graphicx}% Include figure files
\usepackage{dcolumn}% Align table columns on decimal point
\usepackage{bm}% bold math
\begin{document}

\title {Theoretical study of the insulating oxides and nitrides: \\
SiO$_{2}$, GeO$_{2}$, Al$_{2}$O$_{3}$, 
Si$_{3}$N$_{4}$, and Ge$_{3}$N$_{4}$}

\author{C. Sevik}
\email{sevik@fen.bilkent.edu.tr}
\author{C. Bulutay}
\email{bulutay@fen.bilkent.edu.tr}
 \affiliation{Department of Physics and Institute of Materials Science and Nanotechnology\\
  Bilkent University, Ankara, 06800, Turkey}
\date{\today}

\begin{abstract}
An extensive theoretical study is performed for wide bandgap 
crystalline oxides and nitrides, namely, SiO$_{2}$, GeO$_{2}$, 
Al$_{2}$O$_{3}$, Si$_{3}$N$_{4}$, and Ge$_{3}$N$_{4}$. 
Their important polymorphs are considered which are for SiO$_{2}$: 
$\alpha$-quartz, $\alpha$- and $\beta$-cristobalite 
and stishovite, for GeO$_{2}$: $\alpha$-quartz, and rutile,
for Al$_{2}$O$_{3}$: $\alpha$-phase, for Si$_{3}$N$_{4}$ and 
Ge$_{3}$N$_{4}$: $\alpha$- and $\beta$-phases. This work constitutes 
a comprehensive account of both electronic structure 
and the elastic properties of these important insulating 
oxides and nitrides obtained with high accuracy based on density functional 
theory within the local density approximation. 
Two different norm-conserving \textit{ab initio} pseudopotentials have been 
tested which agree in all respects with the only exception arising for the elastic 
properties of rutile GeO$_{2}$. The agreement with experimental values, 
when available, are seen to be highly satisfactory.
The uniformity and the well convergence of this 
approach enables an unbiased assessment of important 
physical parameters within each material and among different insulating oxide and 
nitrides. The computed static electric susceptibilities are observed to display 
a strong correlation with their mass densities. There is a marked discrepancy between 
the considered oxides and nitrides with the latter having sudden increase of 
density of states away from the respective band edges. This is expected to give 
rise to excessive carrier scattering which can practically preclude bulk impact 
ionization process in Si$_{3}$N$_{4}$ and Ge$_{3}$N$_{4}$.
\end{abstract}

% 61.50.Ah 	Theory of crystal structure, crystal symmetry; calculations and modeling
% 62.20.Dc 	Elasticity, elastic constants
% 71.20.-b 	Electron density of states and band structure of crystalline solids
% 71.20.Ps 	Other inorganic compounds
% 77.22.Ch 	Permittivity (dielectric function)

\maketitle

\section{Introduction}
Insulating oxides and nitrides are indispensable materials for diverse 
applications due to their superior mechanical, thermal, chemical and other outstanding 
high temperature properties. 
Furthermore, in the electronic industry these wide band gap materials are being 
considered for alternative gate oxides \cite{robertson} and in the field of 
integrated optics they provide low-loss dielectric waveguides \cite{ay}. 
Recently the subject of wide bandgap oxides and nitrides have gained interest 
within the context of nanocrystals which offer silicon-based technology for 
light emitting devices and semiconductor memories \cite{ossicini}.
These nanocrystals are embedded in an insulating matrix which is usually chosen to
be silica \cite{sput,pecvd1,wang02,serincan}. However, other
wide bandgap materials are also employed such as germania \cite{volodin,gorokhov},
silicon nitride \cite{volodin2,steimle,choi}, and 
alumina \cite{tognini,wan1,tetel}. As a matter of fact, the effect of 
different host matrices is an active research topic in this field.

Among these insulating oxides and nitrides technologically most important 
ones are SiO$_{2}$, Al$_{2}$O$_{3}$, Si$_{3}$N$_{4}$. The activity around 
GeO$_{2}$ is steadily increasing.
Another closely-related material, Ge$_{3}$N$_{4}$ has attracted far less 
attention up to now even though it has certain interesting properties 
\cite{cham}. The major obstacle has been the sample growth. However, 
a very recent study reported an \textit{in situ} Ge$_3$N$_4$ growth on Ge, 
demonstrating high thermal stability and large band offsets with respect 
to the Ge system \cite{sjwang}.
In this comprehensive work, we present the \textit{ab initio} structural and 
electronic properties of all these materials considering their common
polymorphs; these are for
SiO$_{2}$: $\alpha$-quartz, $\alpha$- and $\beta$-cristobalite and
stishovite phases, for GeO$_{2}$: $\alpha$-quartz, and rutile phases,
for Si$_{3}$N$_{4}$ and Ge$_{3}$N$_{4}$: $\alpha$- and $\beta$-phases and for
Al$_{2}$O$_{3}$: $\alpha$-phase. For amorphous and inherently
imperfect matrices, these perfect crystalline phases serve as
important reference systems. Moreover, due to their distinct
advantages, \textit{epitaxial} host lattices are preferred over the 
amorphous ones for specific applications.

With an eye on these technological applications, we focus on several 
physical properties
of these lattices. The  elastic constants play an important role on the
strain profile of the embedded  core semiconductor. Using Eshelby's
continuum elastic consideration \cite{eshelby} the radial and tangential
stress fields of the nanocrystal can be determined \cite{balasub};  these in turn,
affect the optical properties \cite{wang02}. The static  and optical
dielectric constants of these lattices introduce nontrivial local field
effects that modify the absorption spectra of an isolated nanocrystal when 
embedded inside one of these matrices \cite{weiss02}. Based on the
simple effective medium theory which has been tested by
\textit{ab initio} calculations \cite{weiss03}, one can assess which host
lattice and nanocrystal combination would possess the desired optical properties.
Because of the dielectric mismatch between the nanocrystal core and the
surrounding lattice, image charges will be produced \cite{efremov}. These
image charges should be taken into account in characterizing
nanocrystal excitons \cite{brus}. Another promising application is the 
visible and near infrared electroluminescence 
from Si and Ge nanocrystals \cite{ossicini}. 
The electroluminescence is believed to be achieved by the recombination of the 
electron hole pairs injected to nanocrystals under high bias \cite{ossicini}. 
In this context the bulk state impact ionization process which can also give rise 
to electroluminescence is considered to be detrimental leading to 
dielectric breakdown.
\begin{table*}
\caption{\label{structure}Structural information on crystals.}
\begin{ruledtabular}
\begin{tabular}{llllcc}
Crystal&Crystal&Lattice Constants (\AA) &Space Group&Molecules Per &Density\\
&Structure&&&Prim. Cell&(gr/cm$^{3}$)\\
\hline
$\alpha$-quartz SiO$_{2}$ & Hexagonal & $a=$4.883\footnotemark[1] 
4.854\footnotemark[2] 4.913\footnotemark[3]& $P3_{2}21$& 3 & 2.698\\ 
&  &$c=$5.371\footnotemark[1]  5.341\footnotemark[2] 5.405\footnotemark[3]& &
&\\
\hline
$\alpha$-cris. SiO$_{2}$ & Tetragonal  & $a=$4.950\footnotemark[1] 
4.939\footnotemark[2]  4.973\footnotemark[3] & $P4_{1}2_{1}2$& 4 & 2.372\\
 &  &$c=$6.909\footnotemark[1]  6.894\footnotemark[2]  6.926\footnotemark[3]& &
&\\ \hline
$\beta$-cris. SiO$_{2}$ & Cubic  & $a=$7.403\footnotemark[1] 
7.330\footnotemark[2]  7.160\footnotemark[3] & $Fd3m$ & 2 &1.966\\
\hline
Stishovite SiO$_{2}$ & Tetragonal & $a=$4.175\footnotemark[1] 
4.145\footnotemark[2]  4.179\footnotemark[4]& $P4_{2}/mnm$ & 2 & 4.298\\
& & $c=$2.662\footnotemark[1]  2.643\footnotemark[2]  2.665\footnotemark[4]& &
&\\ \hline
$\alpha$-quartz GeO$_{2}$ & Hexagonal & $a=$4.870\footnotemark[1]
4.861\footnotemark[2] 4.984\footnotemark[6]& $P3_{2}21$ & 3 & 4.612\\
& & $c=$5.534\footnotemark[1]  5.520\footnotemark[2]  5.660\footnotemark[6]& &
&\\ \hline
Rutile GeO$_{2}$ & Tetragonal  & $a=$4.283\footnotemark[1] 
4.314\footnotemark[2] 4.4066\footnotemark[7]& $P4_{2}/mnm$ & 2 & 6.655\\
& Tetragonal  & $c=$2.782\footnotemark[1]  2.804\footnotemark[2]
2.8619\footnotemark[7]& & &\\ \hline
$\alpha$-Al$_{2}$O$_{3}$ & Rombohedral  & $a=$4.758\footnotemark[1]  
4.762\footnotemark[5] &$R\overline{3}c$ & 2 & 3.992\\
 &  & $c=$12.98\footnotemark[1]  12.896\footnotemark[5]  & & &\\
\hline
$\alpha$-Si$_{3}$N$_{4}$ & Hexagonal & $a=$7.732\footnotemark[1]
7.766\footnotemark[9]& $C_{3v}^{4}$ & 4 &3.211\\
 &  & $c=$5.603\footnotemark[1] 5.615\footnotemark[9]& & &\\ \hline
$\beta$-Si$_{3}$N$_{4}$ & Hexagonal & $a=$7.580\footnotemark[1]
7.585\footnotemark[10]& $C_{6h}^{2}$ & 2 &3.229\\
&  & $c=$2.899\footnotemark[1] 2.895\footnotemark[10]& & &\\
\hline
$\alpha$-Ge$_{3}$N$_{4}$ & Hexagonal & $a=$7.985$^{\rm a}$& $C_{3v}^{4}$ & 4 &5.691\\
 &  & $c=$5.786$^{\rm a}$& & &\\ \hline
$\beta$-Ge$_{3}$N$_{4}$ & Hexagonal & $a=$7.826$^{\rm a}$& $C_{6h}^{2}$ & 2 &5.727\\
&  & $c=$2.993$^{\rm a}$& & &
\end{tabular}
\end{ruledtabular}
\footnotetext[1]{This Work KA}
\footnotetext[2]{This Work FHI}
\footnotetext[3]{Ref.~\onlinecite{wyckoff}}
\footnotetext[4]{Ref.~\onlinecite{baur}}
\footnotetext[6]{Ref.~\onlinecite{balitski}}
\footnotetext[7]{Ref.~\onlinecite{bolzan, wang}}
\footnotetext[5]{Ref.~\onlinecite{ching94}}
\footnotetext[9]{Ref.~\onlinecite{ching}}
\footnotetext[10]{Ref.~\onlinecite{idrobo}}
\end{table*}
For high-field carrier transport, the crucial 
physical quantity was identified to be the valence and conduction band 
density of states (DOS) for each of the crystalline polymorph \cite{fischetti}.
Based on these technology-driven requirements we compute the elastic constants, 
band structures, dielectric permittivities and electronic DOS of these 
aforementioned crystal polymorphs.
Our \textit{ab initio} framework is based
on the density functional theory \cite{kohn,kohn2}, using pseudopotentials and
a plane wave basis \cite{gonze}. 
With the exception of Ge$_{3}$N$_{4}$ which was far less studied, vast amount 
of theoretical work is already available spread throughout
the literature based on a variety of techniques 
\cite{ching2,ching3,liu,ching94,ching,ching98,demuth,holm,christie,idrobo}.
Our first-principles study here enables a uniform comparison of important physical 
parameters within each material and among different insulating oxides and nitrides.

The plan of the paper is as follows: in Sec.~II we provide details of our
{\em ab initio} computations, Sec.~III contains our first-principles 
results for the structural, electronic properties of the
materials considered followed by our conclusions in Sec.~IV.

\begin{table*}
\caption{\label{bonds}Bond lengths and bond angles (in degrees) of 
SiO$_{2}$  and GeO$_{2}$ polymorphs where x represents a Si or a Ge atom.}
\begin{ruledtabular}
\begin{tabular}{llcccccccc}
Crystal&&
x-O (\AA)&x-O (\AA)&O-x-O&O-x-O&O-x-O&O-x-O&x-O-x&x-O-x\\
\hline 
$\alpha$-quartz SiO$_{2}$& This Work
&1.613&1.618&110.75&109.32&109.07&108.47&140.55&\\ 
& Exp.\footnotemark[1]&1.605&1.614&110.50&109.20&109.00&108.80&143.7&\\ \hline
 $\alpha$-quartz GeO$_{2}$& This
Work &1.693&1.699&113.03&110.62&107.94&106.16&130.56&\\ \hline
$\alpha$-cris. SiO$_{2}$& This
Work &1.597&1.596&111.59&110.08&109.03&108.02&146.02&\\ 
&Exp.\footnotemark[2]&1.603&1.603&111.40&110.00&109.00&108.20&146.5&\\
\hline 
$\beta$-cris. SiO$_{2}$& This Work &1.603&&109.47&&&&180&\\

& Exp.\footnotemark[3]&1.611&&107.80&&&&180.00&\\
\hline 
Stishovite SiO$_{2}$& This Work &1.804&1.758&98.47&81.53&&&130.76&98.47\\

& Exp.\footnotemark[4]&1.760&1.810&&&&&130.60&\\
\hline 
Rutile GeO$_{2}$& This Work &1.848&1.824&99.34&80.66&&&99.34&130.33\\
\end{tabular}
\end{ruledtabular}
\footnotetext[1]{Ref.~\onlinecite{levien}}
\footnotetext[2]{Ref.~\onlinecite{downs}}
\footnotetext[3]{Ref.~\onlinecite{wright}}
\footnotetext[4]{Ref.~\onlinecite{endo}}
\end{table*}
\begin{table*}[t!]
\caption{\label{elastic}Elastic constants and bulk modulus for each crystal.}
\begin{ruledtabular}
\begin{tabular}{llcccccccc}
Crystal&(GPa)&$C_{11}$&$C_{12}$&$C_{13}$&$C_{14}$&$C_{33}$&$C_{44}$
&$C_{66}$&$B$ \\
\hline
$\alpha$-quartz SiO$_{2}$&KA&76.2&11.9&11.2&-17.0&101.7&54.0&32.1 &35\\
&FHI&79.5&9.73&9.54&-18.9&101.7&55.5&34.9 &35\\
&Exp.\footnotemark[1]&87.0&7.00&13.0&-18.0&107.0&57.0&40.0 &38\\
&Exp.\footnotemark[2]&87.0&7.00&19.0&-18.0&106.0&58.0& &40\\
\hline
$\alpha$-Cris. SiO$_{2}$&KA&49.30&5.26&-11.41& &44.78&74.15&26.85&12\\
\hline
$\beta$-Cris. SiO$_{2}$&KA&194.0&135.0& & & &82.67& &155\\
&FHI&196.1&134.2& & & &85.40& &155\\
\hline
Stishovite SiO$_{2}$&KA&447.7&211.0&203.0& &776.0&252.0&302.0 &306\\
&FHI&448.8&211.1&191.0& &752.0&256.5&323.0 &302\\
&Exp.\footnotemark[3]&453.0&211.0&203.0& &776.0&252.0&302.0 &308\\
\hline
$\alpha$-quartz GeO$_{2}$&KA&66.7&24.3&23.1&-3.00&118.7&41.3&21.2 &41\\
&FHI&63.8&25.7&26.2&-0.81&120.2&35.3&19.1 &42\\
&Exp.\footnotemark[4]&66.4&21.3&32.0&-2.20&118.0&36.8&22.5 &42\\
&Exp.\footnotemark[2]&64.0&22.0&32.0&-2.00&118.0&37.0&21.0&42\\
\hline
Rutile GeO$_{2}$&KA&405.9&235.3&189.2& &672.4&206.0&314.4 &292\\
&FHI&349.2&197.2&185.1& &617.5&171.8&274.8 &258\\
&Exp.\footnotemark[5]&337.2&188.2&187.4& &599.4&161.5&258.4 &251\\
\hline
$\alpha$-Al$_{2}$O$_{3}$&KA&493.0&164.1&130.1& &485.8&155.5&164.4 &258\\
&Exp.\footnotemark[6]&497.0&164.0&111.0& &498.0&147.0& &251\\ \hline
$\beta$-Si$_{3}$N$_{4}$&KA&421.8&197.8&116.6 & &550.7 &100.2&112.0 &250\\
&Exp.\footnotemark[7]&433.0&195.0&127.0 & &574.0 &108.0&119.0 &259\\
&Exp.\footnotemark[8]&439.2&181.8&149.9 & &557.0 &114.4&135.9 &265\\
\hline
$\beta$-Ge$_{3}$N$_{4}$&KA&364.3&184.9&111.7 & &486.3 &80.4&89.7&225
\end{tabular}
\end{ruledtabular}
\footnotetext[1]{Ref.~\onlinecite{mcskimin}}
\footnotetext[2]{Ref.~\onlinecite{polian}}
\footnotetext[3]{Ref.~\onlinecite{weidner}}
\footnotetext[4]{Ref.~\onlinecite{balitski}}
\footnotetext[5]{Ref.~\onlinecite{wang}}
\footnotetext[6]{Ref.~\onlinecite{watchman}}
\footnotetext[7]{Ref.~\onlinecite{vogel}}
\footnotetext[8]{Ref.~\onlinecite{wendel}}
\end{table*}

\section{Details of {\it Ab initio} Computations}
Structural and electronic properties of the polymorphs
under consideration have been calculated within the density
functional theory \cite{kohn,kohn2}, using the plane wave basis
pseudopotential method as implemented in the ABINIT code \cite{gonze}.
The results are obtained under the local density approximation (LDA) where
for the exchange-correlation interactions we use the Teter
Pade parameterization \cite{xcteter}, which reproduces Perdew-Zunger \cite{perdew}
 (which reproduces the quantum Monte Carlo electron gas data of Ceperley
and Alder \cite{ceperley}). We tested the results under two
different norm-conserving Troullier and Martins \cite{tm1} type
pseudopotentials, which were generated by  A. Khein and D.C. Allan (KA)
and Fritz Haber Institute (FHI). For both pseudopotentials, the
valence configurations of the constituent atoms were chosen as
N($2s^2p^3$), O($2s^2p^4$), Al($3s^2 3p^1$), Si($3s^2 3p^2$), and
Ge($4s^2 4p^2$).
The number of angular momenta of the KA (FHI) pseudopotentials and the
chosen local channel were respectively, for N: 1, $p$ (3, $d$), for O: 1,
$p$ (3, $d$), for Al: 2, $d$ (3, $d$), for Si: 2, $d$ (3, $d$), and for Ge:
1, $p$ (3, $s$).
Our calculated values for these two types of pseudopotentials were very
similar, the only exceptional case being the elastic constants for
rutile GeO$_{2}$. Dielectric permitivity and the fourth-order tensor of 
elastic constants of each crystal are determined by starting from relaxed
 unit cell under the application of finite deformations within density
 functional perturbation theory \cite{elastik1} as implemented in ABINIT
 and ANADDB extension of it. Another technical detail is related with the
 element and angular momentum-resolved partial density of states (PDOS).
To get a representative PDOS behavior we need to specify the spherical 
regions situated around each relevant atomic site. The radii of these 
spheres are chosen to partition the bond length in proportion to the 
covalent radii of the constituent atoms. This resulted in
the following radii: for the $\alpha$-quartz SiO$_2$, $r_{Si}=0.97$~\AA, 
$r_{O}=0.65$~\AA, for the rutile GeO$_2$,
$r_{Ge}=1.16$~\AA, $r_{O}=0.69$~\AA, for
the $\alpha$-Al$_{2}$O$_3$, $r_{Al}=1.32$~\AA, $r_{O}=0.56$~\AA, and 
for the $\beta$-Si$_3$N$_4$, $r_{Si}=1.03$~\AA, $r_{N}=0.70$~\AA. It should be pointed that even though such an approach presents a good relative weight of the elements
and angular momentum channels, it inevitably underestimates the total
DOS, especially for the conduction bands. Other details of the computations
are deferred to the discussion of each crystal polymorph.

\begin{table}[h!]
\caption{\label{dielectric}Dielectric permittivity tensor.}
\begin{tabular}{lcccc}
\hline \hline 
Crystal&$\epsilon_{xx}^{0}=\epsilon_{yy}^{0}$&$\epsilon_{zz}^{0}$&
$\epsilon_{xx}^{\infty}=\epsilon_{yy}^{\infty}$&$\epsilon_{zz}^{\infty}$\\
\hline
$\alpha$-quartz SiO$_{2}$&4.643&4.847&2.514&2.545\\ \hline
$\alpha$-cris. SiO$_{2}$&4.140&3.938&2.274&2.264 \\ \hline
$\beta$-cris. SiO$_{2}$&3.770&3.770&2.078&2.078\\ \hline
Stishovite SiO$_{2}$&10.877&8.645&3.341&3.510\\ \hline
$\alpha$-quartz GeO$_{2}$&5.424&5.608&2.864&2.947 \\ \hline
Rutile GeO$_{2}$&10.876&8.747& 3.679&3.945\\ \hline
$\alpha$-Al$_{2}$O$_{3}$&10.372&10.372&3.188&3.188\\
\hline
$\beta$-Si$_{3}$N$_{4}$&8.053&8.053&4.211&4.294\\\hline
$\beta$-Ge$_{3}$N$_{4}$&8.702&8.643&4.558&4.667\\
\hline \hline 
\end{tabular}
\end{table}

\section{First-principles Results}
\begin{figure*}[t!]
\includegraphics[width=11.5cm]{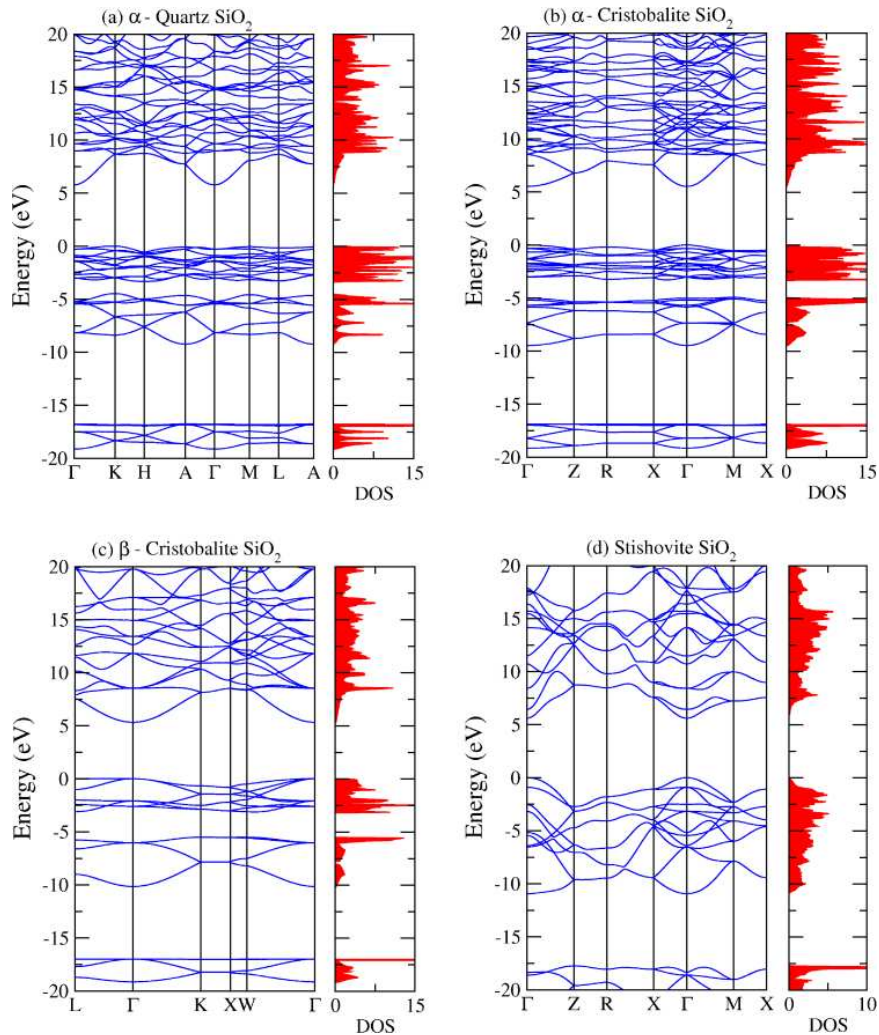}
\caption{\label{BDSiO2}LDA band structure and total DOS (electrons/eV cell) of (a)
$\alpha$-cristobalite SiO$_{2}$, (b) $\alpha$-quartz SiO$_{2}$,
(c) $\beta$-cristobalite SiO$_{2}$, and (d) stishovite SiO$_{2}$. }
\end{figure*}

\begin{figure*}[t!]
\begin{center}
\includegraphics[width=11.5cm]{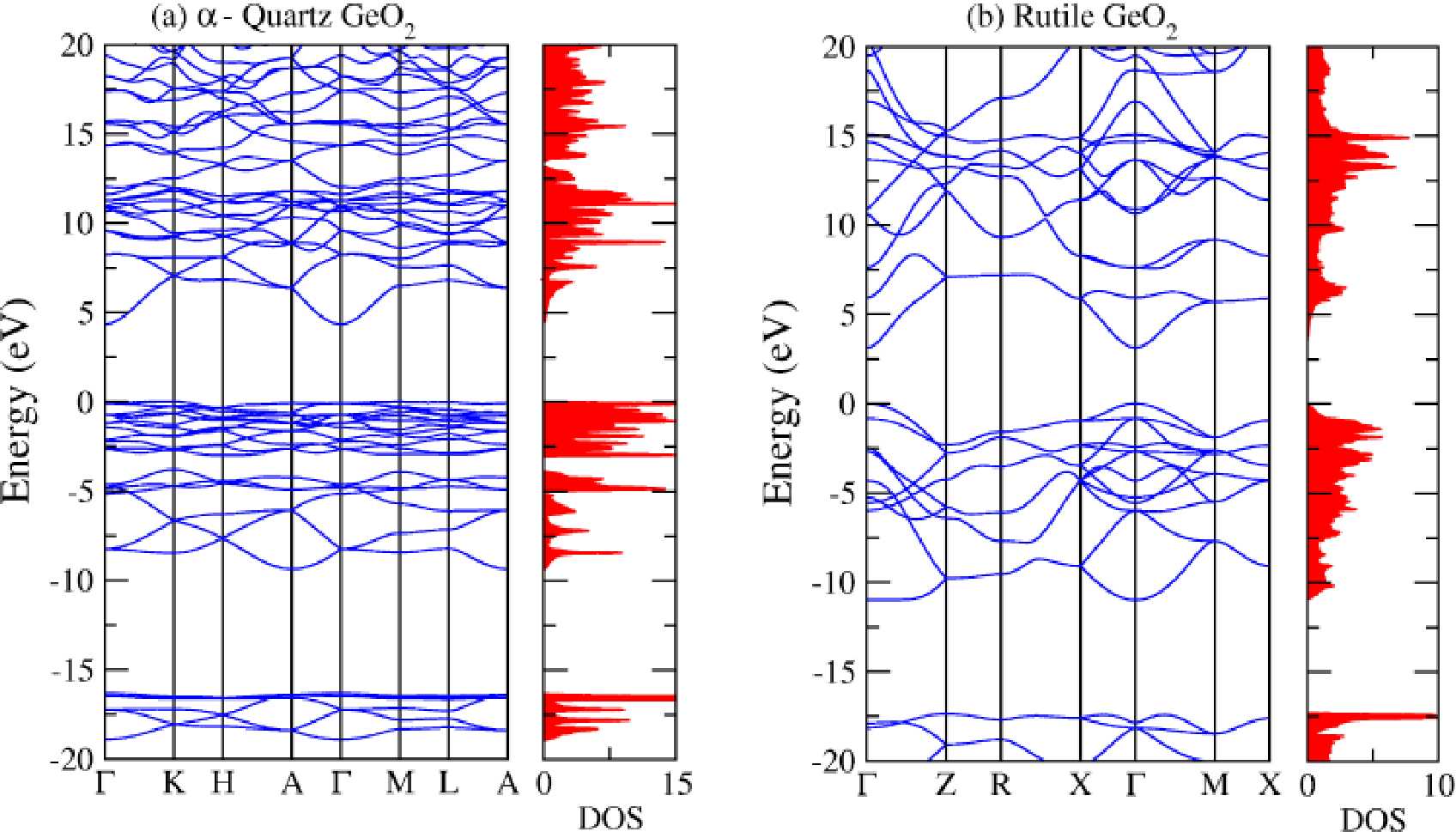}
\end{center}
\caption{\label{BandGeO2}LDA band structure and total DOS of (a) $\alpha$-quartz
GeO$_{2}$, (c) rutile GeO$_{2}$. }
\end{figure*}

\begin{figure}[t!]
\begin{center}
\includegraphics[width=5cm]{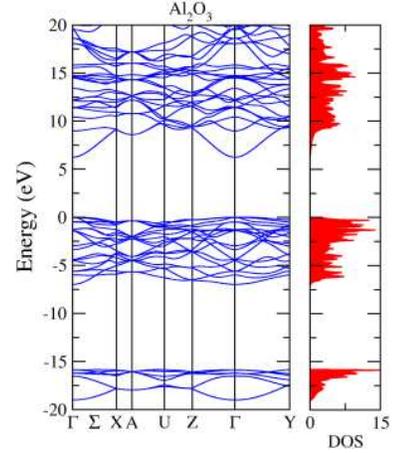}
\end{center}
\caption{\label{BandAl2O3}LDA band structure of and total DOS of $\alpha$-Al$_{2}$O$_{3}$.}
\end{figure}

\begin{figure*}[t!]
\begin{center}
\includegraphics[width=11.5cm]{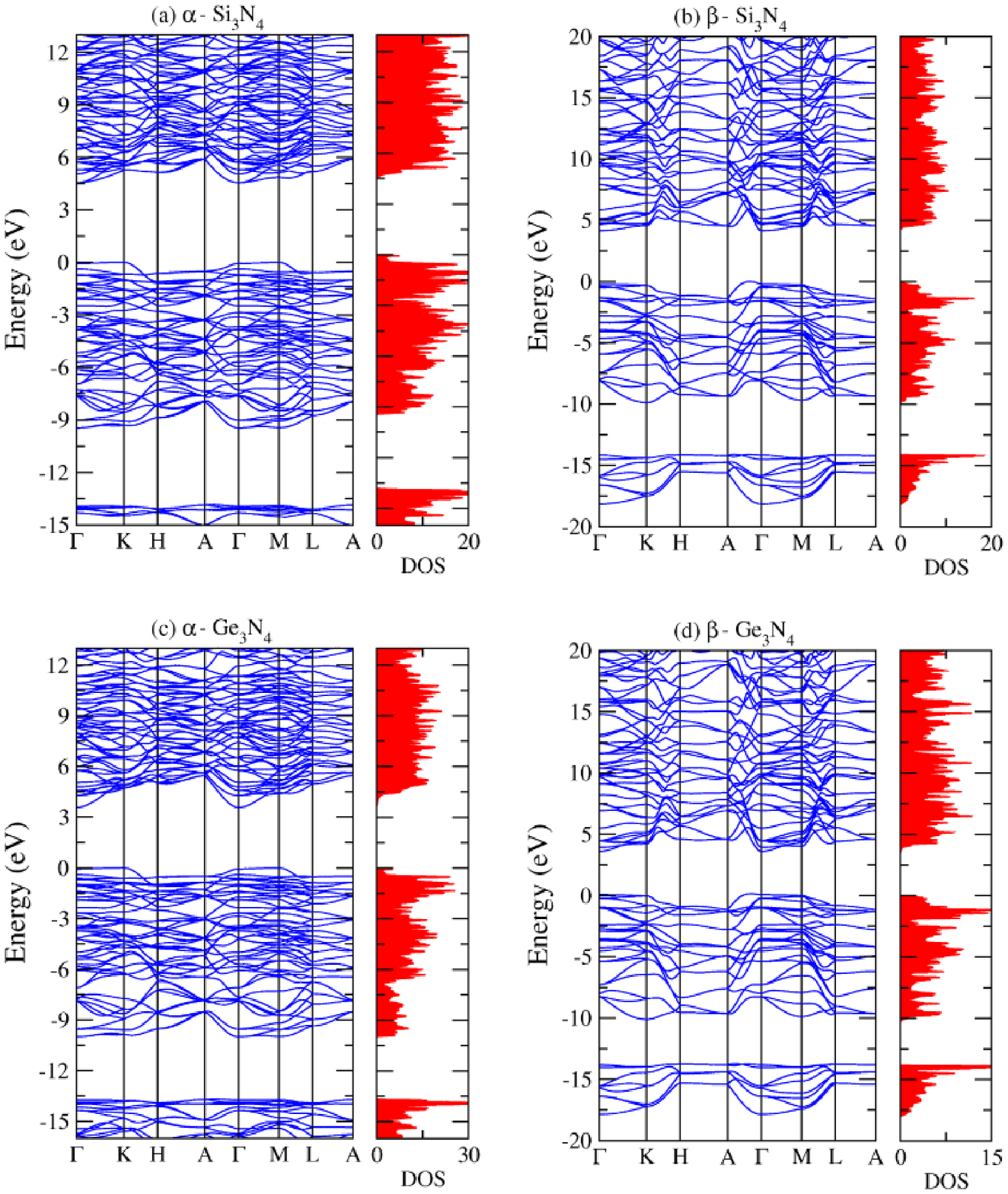}
\end{center}
\caption{\label{BDSiGe3N4}LDA band structure and total DOS of (a) $\alpha$-Si$_{3}$N$_{4}$, (b) $\beta$-Si$_{3}$N$_{4}$, (c) $\alpha$-Ge$_{3}$N$_{4}$ and (d) $\beta$-Ge$_{3}$N$_{4}$.}
\end{figure*}

First, we address the general organization and the underlying 
trends of our results.
The lattice constants and other structural informations of all crystals
are listed in Table~\ref{structure}. Table~\ref{bonds} contains the bond
lengths and bond angles of the optimized oxide polymorphs. These results can
be used to identify the representation of each polymorph within the
amorphous oxides \cite{ginhoven}. The elastic constants and
dielectric permittivity tensor of each crystal are tabulated
in Table~\ref{elastic} and Table~\ref{dielectric}, respectively. Very
close agreement with the existing experimental data and previous
calculations can be observed which gives us confidence about the accuracy 
and convergence of
our work. Employing KA pseudopotentials, the band structure for the crystals
are displayed along the high-symmetry lines
in Figs.~\ref{BDSiO2},\ref{BandGeO2},\ref{BandAl2O3},\ref{BDSiGe3N4} together with
their corresponding total DOS. Such an information is particlulary useful in the 
context of high-field carrier transport.
These results are in good agreement with the
previous computations \cite{ching3,christie,holm,ching}.
\begin{table}[h!]
\caption{\label{egap}Indirect ($E_{g}$) and direct ($E_{g}(\Gamma$)) 
LDA Band Gaps for each crystal.}
\begin{tabular}{lcccc}
\hline \hline
Crystal&VB Max.&CB Min.&$E_{g}$ (eV)&$E_{g}(\Gamma$) (eV)\\
\hline 
$\alpha$-quartz SiO$_{2}$&K&$\Gamma$&5.785&6.073\\ \hline
$\alpha$-cris. SiO$_{2}$&$\Gamma$&$\Gamma$&5.525&5.525\\ \hline
$\beta$-cris. SiO$_{2}$&$\Gamma$&$\Gamma$&5.317&5.317\\ \hline
Stishovite SiO$_{2}$&$\Gamma$&$\Gamma$&5.606&5.606\\ \hline
$\alpha$-quartz GeO$_{2}$&K&$\Gamma$&4.335&4.434\\ \hline
Rutile GeO$_{2}$&$\Gamma$&$\Gamma$&3.126&3.126\\ \hline
$\alpha$-Al$_{2}$O$_{3}$&$\Gamma$&$\Gamma$&6.242&6.242\\\hline
$\alpha$-Si$_{3}$N$_{4}$&M&$\Gamma$&4.559&4.621\\\hline
$\beta$-Si$_{3}$N$_{4}$&A-$\Gamma$&$\Gamma$&4.146&4.365\\\hline
$\alpha$-Ge$_{3}$N$_{4}$&M&$\Gamma$&3.575&3.632\\\hline
$\beta$-Ge$_{3}$N$_{4}$&A-$\Gamma$&$\Gamma$&3.447&3.530\\
\hline \hline 
\end{tabular}
\end{table}
For all of
the considered polymorphs the conduction band minima occur at the
$\Gamma$ point whereas the valence band maxima shift away from this point
for some of the phases making them indirect band gap matrices
(see Table~\ref{egap}). However, the direct band gap values are only
marginally above the indirect band gap values. These LDA band gaps are
underestimated which is a renown artifact
of LDA for semiconductors and insulators \cite{march}. In this work we do
not attempt any correction procedure to adjust the LDA band gap values. 

\begin{figure}[t!]
\begin{center}
\includegraphics[width=8cm]{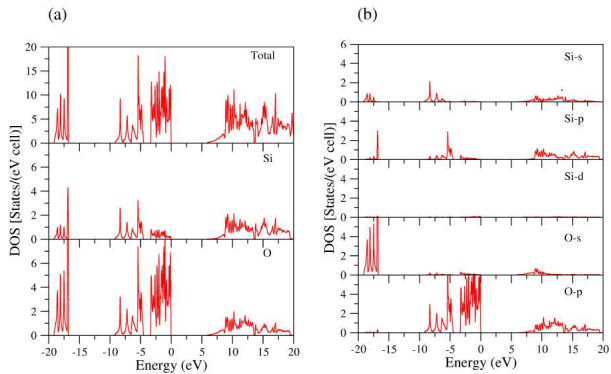}
\end{center}
\caption{\label{AquarDos}DOS of $\alpha$-quartz SiO$_{2}$ (a) Element-resolved;
 total, PDOS of Si, PDOS of O. (b) Angular momentum-resolved; Si $s$ electrons, 
Si $p$ electrons, Si $d$ electrons (not visible at the same scale), O $s$ electrons, O $p$ electrons.}
\end{figure}

\begin{figure}[b!]
\begin{center}
\includegraphics[width=8cm]{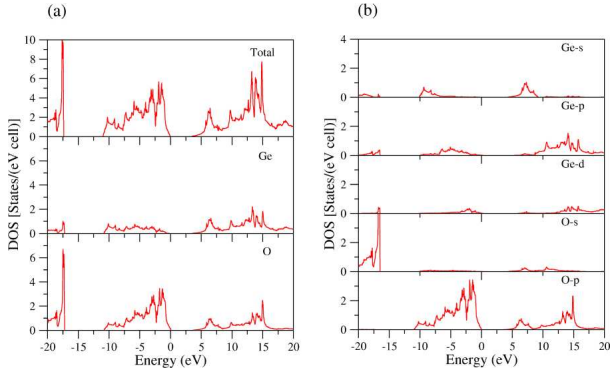}
\end{center}
\caption{\label{RutileDos}DOS of rutile GeO$_{2}$ (a) Element-resolved; total
PDOS of Ge, PDOS of O. (b) Angular momentum-resolved; Ge $s$ electrons, 
Ge $p$ electrons, Ge $d$ electrons, O $s$ electrons, O $p$ electrons.}
\end{figure}

\begin{figure}[ht!]
\begin{center}
\includegraphics[width=8cm]{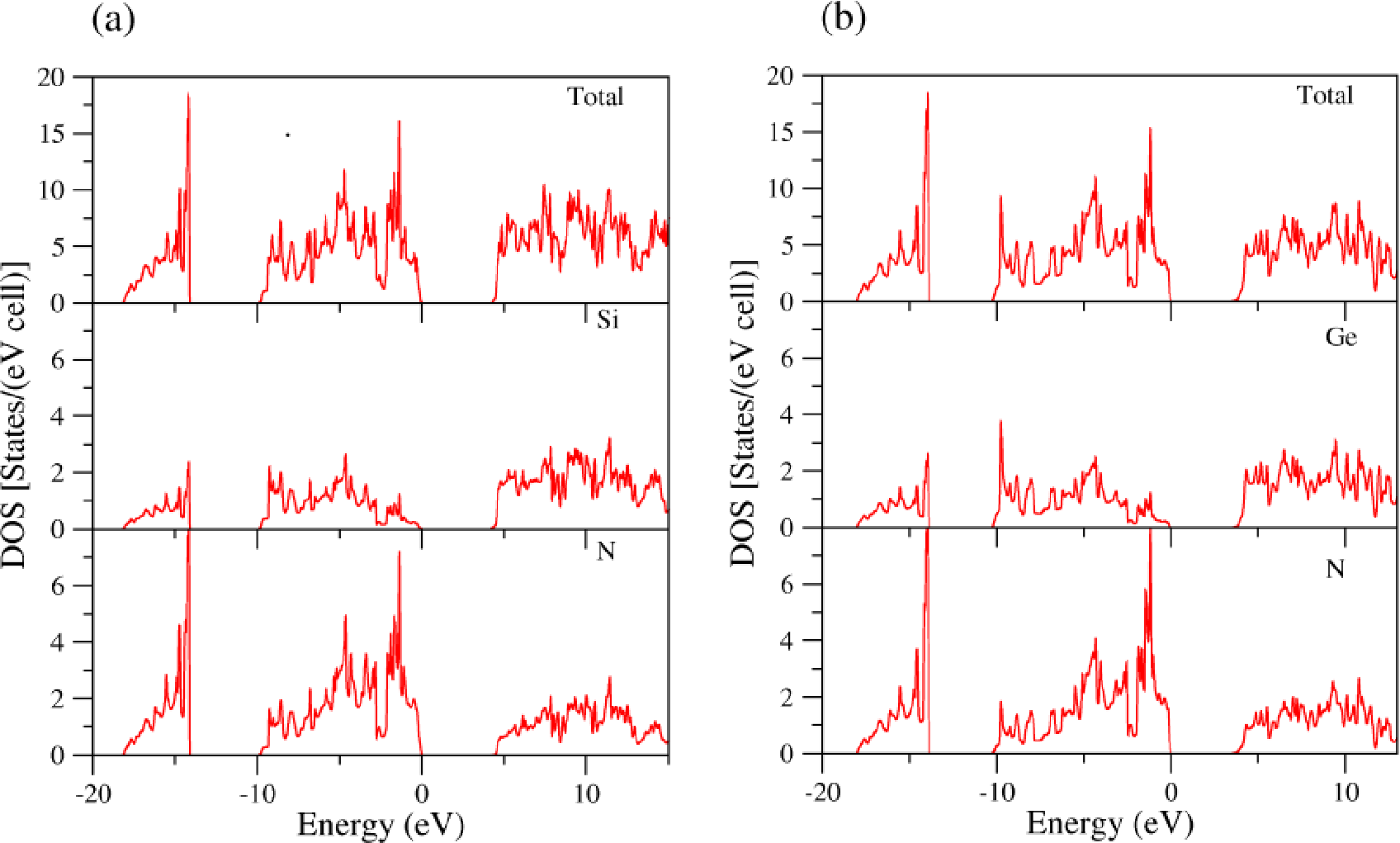}
\end{center}
\caption{\label{SiGe3N4Dos}Element-resolved DOS of (a) $\beta$-Si$_{3}$N$_{4}$;
total, PDOS of Si, PDOS of N, (b) $\beta$-Ge$_{3}$N$_{4}$; total, PDOS of Ge, 
PDOS of N.}
\end{figure}

We present in Figs.~\ref{AquarDos},\ref{RutileDos},\ref{SiGe3N4Dos} 
the element- and angular momentum-resolved PDOS. 
A common trend that can be observed in these various lattices is that
their valence band maxima are dominated by the $p$ states belonging to O
atoms; in the case of Si$_3$N$_4$ and 
Ge$_{3}$N$_{4}$ they are the N atoms. For the conduction band edges,
both constituent elements have comparable contribution. This parallels
the observation in amorphous SiO$_2$ where due to large
electronegativity difference between Si and O, the bonding orbitals have a
large weight on O atoms whereas the lowest conduction band states
with antibonding character have a significant contribution from the
Si atoms \cite{sarnthein}. 

From another perspective, the band structures and the 
associated DOS reveal that there is a marked discrepancy between 
the valence and conduction band edges where for the former there occurs a sharp 
increase of DOS just below the band edge. As the probabilities of most 
scattering processes are directly proportional to DOS \cite{ridley}, 
in the case of high-field carrier transport the electrons should 
encounter far less scatterings and hence gain much higher energy from 
the field compared to holes. In this respect Si$_{3}$N$_{4}$ and 
Ge$_{3}$N$_{4}$ are further different from the others where for both conduction 
and valence bands the DOS dramatically increases (cf. Fig.~\ref{BDSiGe3N4}) so 
that the carriers should suffer from excessive scatterings which 
practically precludes the bulk impact ionization for this material.

Another common trend can be investigated between
the density of each polymorph and the corresponding static permittivity, 
$\epsilon_s$. Such a correlation was put forward by Xu and Ching among the
SiO$_2$ polymorphs \cite{ching3}. We extend this comparison to all
structures considered in this work and rather use $\chi_e = \epsilon_s-1$ which corresponds 
to electric susceptibility. It can be observed from Fig.~\ref{epsdensity} that 
the trend established by SiO$_2$ polymorphs is also followed by $\beta$-Si$_3$N$_4$ 
and $\alpha$-Al$_{2}$O$_3$. On the other hand, Ge-containing 
structures while possessing a similar trend among themselves, display a 
significant shift due to much higher mass of the this atom.
This dependence on the atomic mass needs to be removed by finding a
more suitable physical quantity. We should mention that such a correlation
does not exist between the volume per primitive cell of each phase and
the static permittivity. After these general comments, now we concentrate on
the results of each lattice individually.

\subsection{SiO$_{2}$}
% $\alpha$-quartz SiO$_{2}$
The $\alpha$-quartz SiO$_{2}$ is one of the most studied polymorphs as it is
the stable phase at the ambient pressure and
temperature \cite{liu,demuth}, furthermore its short-range order is
essentially the same as the amorphous SiO$_{2}$ \cite{sarnthein}.
$\alpha$-quartz SiO$_{2}$ has a hexagonal unit cell containing three
SiO$_{2}$ molecules. A plane-wave basis set with an energy cutoff of 60~Ha
was used to expand the electronic wave functions at the special $k$-point
mesh generated by 10$\times$10$\times$8 Monkhorst-Pack
scheme \cite{monkhorts}. The band structure of $\alpha$-quartz SiO$_{2}$ has
been calculated by many authors (see, for instance \cite{ching2,ching3}).
Our calculated band structure and total DOS shown in
Fig.~\ref{BDSiO2}(a) are in agreement with the published studies 
\cite{ching3}. The indirect
LDA band gap for this crystal is 5.785~eV from the valence band maximum at
$K$ to the conduction band minimum at $\Gamma$. The direct LDA band gap
at $\Gamma$ is slightly larger than the indirect LDA band gap as seen
in Table~\ref{egap}. 
Calculated values of the elastic constants and bulk modulus listed in
Table~\ref{elastic} are in good agreement with the experiments. Apart
from $C_{12}$, the elastic constants are within 10$\%$ of the
experimental values. The discrepancy in $C_{12}$ can be explained by the
fact that $C_{12}$ is very soft and this type of deviation also exists
among experiments which is also the case for $C_{14}$.

% $\alpha$-Cristobalite SiO$_{2}$
$\alpha$-cristobalite SiO$_{2}$ has a tetragonal unit cell containing
four SiO$_{2}$ molecules. In the course of calculations an absolute
energy convergence of 10$^{-4}$~Ha was obtained by setting a high plane
wave energy cutoff as 60~Ha and 10$\times$10$\times$8 $k$-point
sampling. Figure~\ref{BDSiO2}(b) shows the band structure
of $\alpha$-cristobalite SiO$_{2}$ with the 5.525~eV direct band gap
at $\Gamma$. The bulk modulus of 12~GPa is the smallest among all the 
host lattice polymorphs considered in this work.

% $\beta$-Cristobalite SiO$_{2}$
Regarding $\beta$-cristobalite, its actual structure is somewhat controversial, as
a number of different symmetries have been proposed corresponding to
space groups $Fd3m$, $I\overline{4}2d$, and $P2_{1}3$ \cite{demuth}.
Recently, incorporating the quasiparticle corrections the
tetragonal $I\overline{4}2d$ phase was identified to be energetically
most stable \cite{ramos}. However, we work with the structure having the
space group of $Fd3m$ that was originally proposed by Wyckoff \cite{wyck1925}
and which is widely studied primarily due to
its simplicity \cite{ching2,liu}. This phase has a cubic conventional
cell with two molecules. We used 60~Ha plane wave energy cutoff
and 10$\times$10$\times$10 $k$-point sampling. Figure~\ref{BDSiO2}(c)
shows the band structure of $\beta$-cristobalite SiO$_{2}$ with the
5.317~eV direct band gap at $\Gamma$. Unlike their band structures, total DOS
of $\alpha$- and and $\beta$-cristobalite SiO$_{2}$ are very similar
(cf. Fig.~\ref{BDSiO2}(c)). This similarity can be explained by the fact
that their local structure are very close. On the other hand there is
a considerable difference between the DOS spectra of the
$\alpha$-quartz SiO$_{2}$ and the $\beta$-cristobalite SiO$_{2}$.
In Table~\ref{elastic}, we present elastic constants of the
$\beta$-cristobalite SiO$_{2}$ calculated by two types of pseudopotentials, 
FHI and KA. There is no considerable difference between them. Dielectric 
constants of $\beta$-cristobalite SiO$_{2}$ are the smallest among the 
five polymorphs of SiO$_{2}$ studied here (see Table~\ref{dielectric}).

% Stishovite SiO$_{2}$
Stishovite is a dense polymorph of SiO$_{2}$ with octahedrally coordinated
silicon, unlike the previous phases \cite{demuth}. It has a tetragonal cell
with two molecules. Calculations were done by using 60~Ha plane wave
energy cutoff and 8$\times$8$\times$10 $k$-point sampling. The band structure
of stishovite with a wide single valence band is markedly different from that
of the previous three crystalline phases of SiO$_{2}$ having two narrow
upper valence bands. The cause of this increased valence bandwidth  is the
lack of separation between bonding and nonbonding states \cite{christie}.
 Hence, the total DOS for stishovite shows no gap at the middle of the
valence band (see Fig.~\ref{BDSiO2}(d)). Our calculations yield a direct
LDA band gap of 5.606~eV at $\Gamma$. As seen in Table~\ref{elastic},
the differences between our computed elastic constants and the
experimental values are less than 3$\%$; this is an excellent agreement for
LDA. Its bulk modulus is the largest among all the host lattice
polymorphs considered in this work. Moreover, dielectric constants of
stishovite is the largest of the five polymorphs of SiO$_{2}$ considered in
this work (see Table~\ref{dielectric}).

\subsection{GeO$_{2}$}
% $\alpha$-Quartz GeO$_{2}$
For $\alpha$-quartz GeO$_{2}$ we used the same energy cutoff and $k$-point 
sampling as with $\alpha$-quartz SiO$_{2}$ which yields excellent convergence. 
The band structure of the $\alpha$-quartz GeO$_{2}$ is displayed
in Fig.~\ref{BandGeO2}(a). The similarity of the band structures of
the $\alpha$-quartz GeO$_{2}$ and the $\alpha$-quartz SiO$_{2}$ is
not surprising as they are isostructural. Similarly their total DOS
resemble each other (cf.~Fig.~\ref{BandGeO2}(a)). The indirect LDA band gap
for this phase is 4.335~eV from the valence band maximum at $K$ to
the conduction band minimum at $\Gamma$. The direct band gap at $\Gamma$
is slightly different from indirect band gap as seen in Table~\ref{egap}.
This gap is smaller than that of the $\alpha$-quartz SiO$_{2}$. The
perfect agreement between calculated elastic constants of the
$\alpha$-quartz GeO$_{2}$ and experimental values \cite{polian,balitski} can
be observed in Table~\ref{elastic}.

% Rutile GeO$_{2}$
The rutile structure of GeO$_{2}$, also known as argutite \cite{mineral}
is isostructural with the stishovite phase of SiO$_{2}$. The
same energy cutoff and $k$-point sampling values as for stishovite 
yield excellent convergence. The direct LDA band gap at $\Gamma$ for 
rutile-GeO$_{2}$ is less than
that of stishovite with a value of 3.126~eV. The two upper valence bands
are merged in the total DOS (see Fig.~\ref{BandGeO2}(b)) as in the case
of stishovite. The increased valence bandwidth in the band structure can
be explained by the same reason as in the case of stishovite. The results 
of the elastic constants calculated with KA type pseudopotential shown in 
Table~\ref{elastic} deviate substantially from the experiment whereas
the agreement with the FHI pseudopotentials is highly satisfactory.
The similarity of the dielectric constants of rutile GeO$_{2}$ and
stishovite can be observed in Table~\ref{dielectric}.

\subsection{Al$_{2}$O$_{3}$}
Al$_{2}$O$_{3}$ is regarded as a technologically important oxide due to its
high dielectric constant and being reasonably a good glass former
after SiO$_{2}$ \cite{robertson}. The $\alpha$-Al$_{2}$O$_{3}$ (sapphire) 
has the rhombohedral cell with two molecules. Computations about Al$_{2}$O$_{3}$
were done by using 60~Ha plane wave energy cutoff and a total of 60
$k$-points within the Brillouin zone. Fig.~\ref{BandAl2O3}
shows the computed band structure and total DOS of
the $\alpha$-Al$_{2}$O$_{3}$. These are in excellent agreement with the
previous calculation \cite{ching94,ching98}. For Al$_{2}$O$_{3}$, minimum of
the conduction band is at $\Gamma$ and maximum of the valence band is at a
point along $\Gamma-X$ close to the $\Gamma$ point. The corresponding LDA
band gap is 6.242~eV. Because of the very small difference between the
direct and indirect band gaps, Al$_{2}$O$_{3}$ is considered as a direct
band gap insulator. Measured band gap of this crystal is 8.7~eV. However
the precise value of the gap of Al$_{2}$O$_{3}$ is still elusive because of
the existence of an excitonic peak near the absorbtions edge \cite{french}.
As seen in Table~\ref{elastic}, computed values of the elastic constant and
bulk modulus of  Al$_{2}$O$_{3}$ are in excellent agreement with
the experiments. As a furher remark, the $\alpha$-Al$_{2}$O$_{3}$ unit cell can be described as hexagonal or rhombohedral depending on the crystallographical definition of the space group R$\overline{3}$C. During our first-principles calculations it has been defined as rhombohedral in which case $C_{14}$ vanishes. Although the sign of $C_{14}$ is experimentally determined to be negative for the hexagonal-Al$_{2}$O$_{3}$, previous calculations reported a positive value \cite{page}. To check this disagreement we have calculated the elastic constant of the hexagonal-Al$_{2}$O$_{3}$ and found it to be around -3.0.

\subsection{Si$_{3}$N$_{4}$ and Ge$_{3}$N$_{4}$}
The research on silicon nitride has largely been driven by its use in microelectronics
technology to utilize it as an effective insulating material and also as diffusion 
mask for impurities. Recently it started to attract attention both as a
host embedding material for nanocrystals \cite{volodin2,steimle,choi} and also
for optical waveguide applications \cite{ay}. The $\alpha$-
and $\beta$-Si$_{3}$N$_{4}$ have hexagonal conventional cells with four and
two molecules, respectively. We used 60~Ha plane wave energy cutoff
and 6$\times$6$\times$8 $k$-point sampling.
The computed band structures of these two phases shown in Figs.~\ref{BDSiGe3N4}
(a) and (b) are identical to those reported by Xu and Ching \cite{ching}. 
The top of the valence band for $\beta$-Si$_{3}$N$_{4}$ is along the 
$\Gamma$-A direction, and for $\alpha$-Si$_{3}$N$_{4}$ it is at the $M$ point. 
The bottom of the conduction band for two phases are at the $\Gamma$ point.
The direct and indirect LDA band gaps of these two phases are
respectively, 4.559~eV, 4.621~eV for  $\alpha$-Si$_{3}$N$_{4}$ and 
4.146~eV, 4.365~eV for the $\beta$-Si$_{3}$N$_{4}$. 
The general band structure of two phases
are very similar, except that the $\alpha$-Si$_{3}$N$_{4}$ has twice as
many bands because the unit cell is twice as large. The total DOS of these
two phases shown in Figs.~\ref{BDSiGe3N4}(a) and (b) are only
marginally different. Calculated values of the elastic constants and
bulk modulus of $\beta$-Si$_{3}$N$_{4}$ listed in Table~\ref{elastic} are
in excellent agreement with the quoted experiments. Those for
the $\alpha$-Si$_{3}$N$_{4}$ which is thermodynamically less stable with
respect to $\beta$-phase \cite{weiss81} were left out due to excessive
memory requirements for the desired accuracy.

\begin{figure}[ht!]
\begin{center}
\includegraphics[width=8cm]{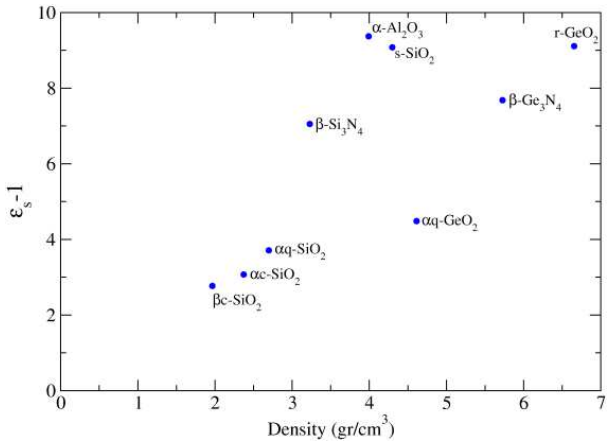}
\end{center}
\caption{\label{epsdensity}Density versus direction-averaged
static electric susceptibility.}
\end{figure}

Ge$_{3}$N$_{4}$ is the least studied material among the oxides and nitrides 
considered in this work. Recently its high-pressure $\gamma$-phase has attracted 
some theoretical interest \cite{dong}. However, the available Ge$_{3}$N$_{4}$ samples 
contain a mixture of $\alpha$ and $\beta$-phases as in the case of Si$_{3}$N$_{4}$ 
and these are the polymorphs that we discuss in this work. 
The band structures of both of these phases of Ge$_{3}$N$_{4}$ (cf. Fig.~\ref{BDSiGe3N4}) 
are very similar to those of Si$_{3}$N$_{4}$. Regarding the 
elastic constants of $\beta$-Ge$_{3}$N$_{4}$, our theoretical results listed in 
Table~\ref{elastic} await experimental verification. In terms of density, the $\beta$ phases 
of Si$_{3}$N$_{4}$ and Ge$_{3}$N$_{4}$ fill the gap between the $\alpha$-quartz and 
stishovite/rutile phases of their oxides. As can be observed from Fig.~\ref{epsdensity} 
their electric susceptibility versus density behavior strengthens the correlation 
established by the remaining polymorphs. Finally it should be pointed that 
$\beta$-Ge$_3$N$_4$ has the largest high-frequency dielectric constant ($\epsilon_\infty$) 
among all the materials considered in this work.

\section{Conclusions}
A comprehensive first-principles study is presented which is unique in
analyzing common polymorphs of the technologically-important insulating 
oxides and nitrides: SiO$_{2}$, GeO$_{2}$, Al$_{2}$O$_{3}$, Si$_{3}$N$_{4}$, 
and Ge$_{3}$N$_{4}$. The
structural parameters, elastic constants, static and optical
dielectric constants are obtained in close agreement with the
available results. 
The computed dielectric constants are observed to display a strong 
correlation with their mass densities.
For all of the considered polymorphs the conduction
band minima occur at the $\Gamma$ point whereas the valence band
maxima shift away from this point for some of the phases making them
indirect band gap matrices. However, the direct band gap values are
only marginally above the indirect band gap values. 
The investigation of band structure and DOS data reveal that the
holes in all polymorphs considered and the electrons for the case of 
Si$_{3}$N$_{4}$ and Ge$_{3}$N$_{4}$ should suffer excessive scatterings 
under high applied 
field which will preclude bulk impact ionization for these carrier types and 
polymorphs. This can be especially important for applications vulnerable 
to dielectric breakdown.

\subsection{Acknowledgments}
This work has been supported by the European FP6 Project SEMINANO with
the contract number NMP4 CT2004 505285. We would like to thank R. Eryi{\u g}it,
T. G{\"u}rel, O. G\"ulseren, D. {\c C}ak{\i}r and T. Y{\i}ld{\i}r{\i}m for
their useful advices and to Dr. Can U\u{g}ur Ayfer for the access to Bilkent University Computer Center facilities.

\end{document}